\documentclass{PoS}
\newcommand{\be}{\begin{equation}}
\newcommand{\ee}{\end{equation}}
\newcommand{\beq}{\begin{eqnarray}}
\newcommand{\eeq}{\end{eqnarray}}

\usepackage{epsfig}
\usepackage{epsf}
 \usepackage{dsfont}
\usepackage{slashed}
\usepackage{booktabs}
\usepackage{graphicx}
\usepackage{float}

\newcommand{\Op}{\mathcal{O}} 
\newcommand{\C}{\mathcal{C}} 
\newcommand{\eins}{\mathds{1}} 

\title{Nucleon form factors with $N_F=2$ twisted mass fermions}

\ShortTitle{Nucleon form factors with dynamical TMF}

\author{\speaker{C. Alexandrou}~$^{(a,b)}$, T. Korzec\footnote{Current address:Institut f\"ur Physik
   Humboldt Universit\"at zu Berlin, Newtonstrasse 15, 12489 Berlin, Germany.}~$^{(a)}$,
 G. Koutsou~$^{(a,c)}$\\
$^{(a)}$ Department of Physics, University of Cyprus, P.O. Box 20537,
 1678 Nicosia, Cyprus \\
$^{(b)}$  Computation-based Science and Technology Research Center,  Cyprus Institute, 15 Kypranoros St., 1645 Nicosia, Cyprus \\
$^{(c)}$ Bergische Universit\"at Wuppertal, Fachbereich Physik, 42097 Wuppertal, Germany and\\
JSC and IAS, FZ J\"ulich, 52425 J\"ulich, Germany\\
        E-mail: \email{alexand@ucy.ac.cy}, \email{korzec@ucy.ac.cy}, \email{i.koutsou@fz-juelich.de}}
\author{R.~Baron,  P.~Guichon\\
CEA-Saclay, IRFU/Service de Physique Nucl\'eaire, 91191 Gif-sur-Yvette, France\\ 
 E-mail: \email{remi.baron@cea.fr}, \email{pierre.guichon@cea.fr}}
\author{M.~Brinet, J.~Carbonell,   P.-A.~Harraud\\
Laboratoire de Physique Subatomique et Cosmologie,
               UJF/CNRS/IN2P3, 53 avenue des Martyrs, 38026 Grenoble, France\\
 E-mail: \email{mariane@lpsc.in2p3.fr}, \email{Jaume.Carbonell@lpsc.in2p3.fr}, \email{harraud@lpsc.in2p3.fr}}
\author{K. Jansen,\\
 NIC, DESY, Platanenallee 6, D-15738 Zeuthen, Germany\\
Email: \email{Karl.Jansen@desy.de}}
\author{\center{\mbox{\includegraphics[width=0.16\linewidth]{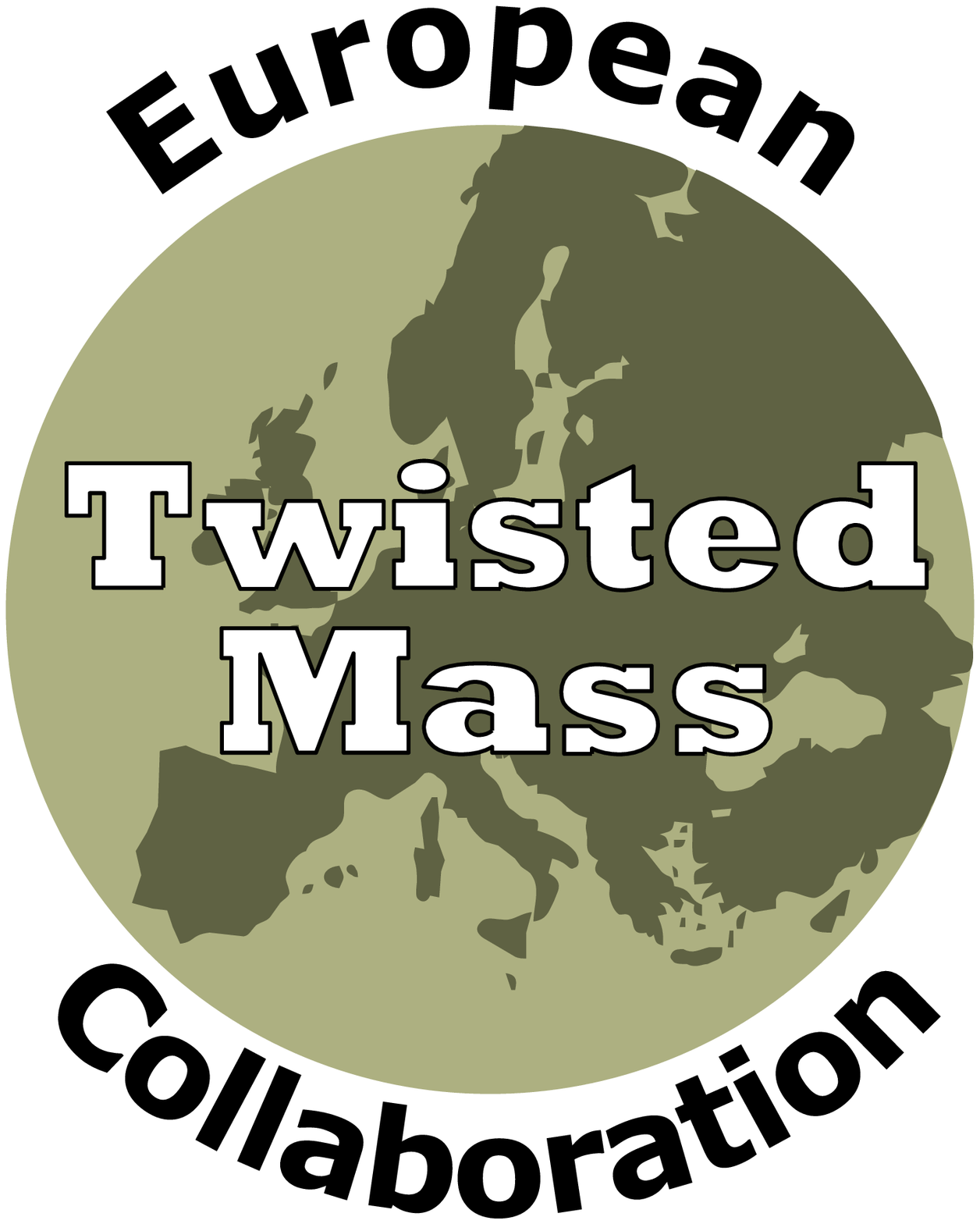}}}}

\abstract{We present results on the electromagnetic and axial nucleon form factors 
using two degenerate flavors of twisted mass fermions
on lattices of spatial size 2.1 fm
and 2.7 fm  and a lattice spacing of about 0.09 fm.
We consider  pion masses in the range of 260-470 MeV. We chirally extrapolate results on the nucleon axial charge, the  isovector Dirac and Pauli root mean squared radii and magnetic moment to the physical point and compare to
experiment. }

\FullConference{The XXVII International Symposium on Lattice Field Theory - LAT2009\\
		 July 26-31 2009\\
		 Peking University, Beijing, China}

\begin{document}

\section{Introduction}
Twisted mass fermions~\cite{Alpha}  provide an attractive  formulation of lattice QCD that
allows for automatic ${\cal O}(a)$ improvement, infrared regularization 
of small
eigenvalues and fast dynamical 
simulations. A particularly attractive feature for the
calculation of the nucleon form factors discussed in this work 
is the automatic  
 ${\cal O}(a)$ improvement obtained by tuning only one parameter,
requiring no further improvements on the operator level.
Important physical results are emerging using gauge configurations
 generated with two degenerate flavors of
twisted quarks ($N_F=2)$ in both the meson~\cite{ETMC-mesons} and baryon~\cite{ETMC-baryons}
sectors. An example is the 
accurate determination, using precise results in the meson sector, of low energy constants
of great relevance to phenomenology. 
 Currently, $N_F=2$ simulations are available for pion mass in the range of about
 260-470~MeV for three lattice spacings $a<0.1$~fm
allowing for
continuum and chiral extrapolations. In this work we 
discuss high-statistics results on the nucleon form factors obtained at 
one value of the lattice spacing.
Electromagnetic and axial form factors (FFs) of the proton 
and the neutron are fundamental quantities
that yield information on their internal structure 
such as their size, magnetization and axial
charge. They have been studied experimentally for a long time
 improving their measuements both in terms of precision as well as 
in terms of probing larger momentum transfers.
Several lattice collaborations are currently using dynamical 
fermions  to calculate these fundamental quantities~\cite{{Ohta:2008kd}, Syritsyn:2009mx}.

The action for two degenerate flavors of quarks
 in twisted mass QCD is given by
   \begin{equation}
      S = S_g + a^4 \sum_x \bar\chi(x) \left[\frac{1}{2}\gamma_\mu(\nabla_\mu + \nabla_\mu^*)  -\frac{ar}{2}\nabla_\mu\nabla_\mu^* 
					    + m_{\rm crit}
					    + i\gamma_5\tau^3\mu \right]\chi(x)\quad ,
   \end{equation}
where  we use the  tree-level Symanzik improved
gauge action $S_g$. The quark fields $\chi$
are in the so-called "twisted basis" obtained from the "physical basis"
at
maximal twist by the transformation
$    \psi = \frac{1}{\sqrt{2}}[{\bf 1} + i\tau^3\gamma_5]\chi$ and
$\bar\psi = \bar\chi \frac{1}{\sqrt{2}}[{\bf 1} + i\tau^3\gamma_5]$.
We note that, in the continuum, this action
 is   equivalent to QCD.
A crucial advantage  is
the fact that by tuning a single parameter, namely the bare untwisted quark mass to its critical value
 $m_{\rm cr}$, physical observables are automatically 
${\cal O}(a)$ improved. 
A disadvantage is  the explicit flavor symmetry breaking. In a recent paper 
we have checked that this breaking is small for baryon observables for
the lattice spacing discussed
here~\cite{Alexandrou:2009}.\vspace*{0.2cm}
\begin{minipage}{0.7\linewidth}
\hspace*{0.5cm}{ To extract the nucleon FFs we need to evaluate the 
nucleon matrix elements $\langle N(p_f,s_f) | {j_\mu} | N(p_i,s_i) \rangle$, where
$|N(p_f,s_f)\rangle$, $|N(p_i,s_i)\rangle$ are nucleon states with final (initial) momentum $p_f (p_i)$ and spin $s_f (s_i)$ and ${j_\mu}$ is either
the electromagnetic current $V^{EM}_\mu(x) = \frac{2}{3} \bar u(x) \gamma_\mu u(x) - \frac{1}{3} \bar d(x) \gamma_\mu d(x)$ or the axial current $A_\mu^a(x) = \bar\psi(x) \gamma_\mu \gamma_5 \frac{\tau^a}{2} \psi(x)$. Whereas the
matrix element of the axial current receives contributions only from the
connected diagram shown in Fig.~1 the electromagnetic one has, in addition,
disconnected contributions. In the isospin limit 
the matrix element of the  isovector electromagnetic 
current $V_\mu^a(x) = \bar\psi(x) \gamma_\mu \frac{\tau^a}{2} \psi(x)$ has
no disconnected contributions~\cite{Alexandrou:2008rp}. Therefore in this work we only evaluate the 
isovector nucleon FFs obtained from the
connected diagram.}
\hfill
  \end{minipage} \begin{minipage}{0.29\linewidth}\vspace{-0.5cm}
     \hspace*{1cm}\includegraphics[width=0.8\linewidth]{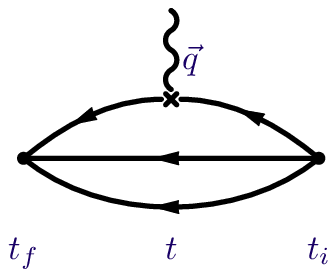}
{\hspace*{0.3cm} Fig.1: Connected nucleon \hspace*{0.3cm} three-point function.}
  \end{minipage}


\vspace*{0.2cm}

  The electromagnetic matrix element of the nucleon can be expressed in terms of the Dirac and Pauli form factors, $F_1$ and $F_2$ defined, in Euclidean time, as \\
$$
     \langle N(p_f,s_f) | V_\mu(0) | N(p_i,s_i) \rangle = \sqrt{\frac{m_N^2}{E_N(\vec p_f) E_N(\vec p_i)}}   \bar u(p_f,s_f) \Op_\mu u(p_i,s_i), \hspace*{0.1cm}
     \Op_\mu = \gamma_\mu { F_1(Q^2)} + \frac{i \sigma_{\mu\nu} q^\nu}{2 m_N} {F_2(Q^2)}
 $$
 with  $q = p_f - p_i$ the momentum transfer and $Q^2=-q^2$.  These are related to the Sachs 
electric $G_E$ and magnetic $G_M$ FFs via:
  $
     {G_E(Q^2)} = {F_1(Q^2)} - \frac{Q^2}{(2m_N)^2}{F_2(Q^2)} $ and
$     {G_M(Q^2)} = {F_1(Q^2)} + {F_2(Q^2)}.$

  Similarly, the axial current matrix element of the nucleon  $ \langle N(p_f,s_f) | A^a_\mu(0) | N(p_i,s_i) \rangle$ can be expressed in terms of the form factors
  $ G_A$ and $G_p$ with ${\cal O}_\mu$ given by
$$
   \Op_\mu = \left[ -\gamma_\mu\gamma_5 { G_A(Q^2)} + i\frac{q^\mu\gamma_5}{2 m_N} {G_p(Q^2)} \right] \frac{\tau^a}{2}.
 $$

\section{Lattice evaluation} 

 The nucleon interpolating field in the physical basis
$       J(x) = \epsilon^{abc} \left[u^{a \top}(x) \C\gamma_5 d^b(x)\right] u^c(x)
$ can be written in the twisted basis  at maximal twist as
$
       \tilde{J}(x) = {\frac{1}{\sqrt{2}}[\eins + i\gamma_5]}\epsilon^{abc} \left[ {\tilde{u}}^{a \top}(x) \C\gamma_5 \tilde{d}^b(x)\right] {\tilde{u}}^c(x).$
The transformation of the electromagnetic current, $V_\mu^a(x)$,  to the twisted basis leaves
the form of   
       $V_\mu^{0,3}(x)$ unchanged.
We use the Noether lattice current and therefore the renormalization
constant $Z_V = 1$.
The  axial current $A_\mu^3 $ also has the same form in the two bases.
 In this case we use the local current and therefore we need 
the renormalization constant $Z_A$.
 The  value of $Z_A=0.76(1)$~\cite{Dimopoulos} was determined non-perturbatively in the RI'-MOM scheme. This value is consistent
with a recent analysis~\cite{Cyprus}, which
uses a perturbative subtraction of 
${\cal O}(a^2)$ terms~\cite{Martha} for a better 
identification of the plateau yielding a value of $Z_A=0.768(3)$~\cite{Cyprus}.
  In order to increase overlap with the proton state and  decrease overlap with excited states
we use Gaussian smeared quark fields~\cite{smearing} for the construction of the interpolating fields:
   $
      {\bf q}^a(t,\vec x) = \sum_{\vec y} F^{ab}(\vec x,\vec y;U(t))\ q^b(t,\vec y) $ with 
 $     F = (\eins + {\alpha} H)^{n} $ and 
 $ H(\vec x,\vec y; U(t)) = \sum_{i=1}^3[U_i(x) \delta_{x,y-\hat\imath} + U_i^\dagger(x-\hat\imath) \delta_{x,y+\hat\imath}].$
  In addition we apply APE-smearing to the gauge fields $U_\mu$ entering $H$.
 The smearing  is the same as for our calculation of baryon masses
 with the smearing parameters $\alpha$ and $n$ optimized for the nucleon
ground state~\cite{ETMC-baryons}.

To set the scale we use the nucleon mass in the physical limit. We show in Fig.~2 
results at three values of the lattice spacings corresponding to $\beta=3.9$,
$\beta=4.05$ and $\beta=4.2$. As can be seen,
cut-off effects are negligible and we can therefore
 use continuum chiral perturbation theory to extrapolate to the physical point.
We correct  for volume dependence coming from pions propagating around the lattice~\cite{QCDSF_vol}. 
To chirally extrapolate we use the well-established ${\cal O}(p^3)$ result of
heavy baryon chiral perturbation theory (HB$\chi$PT) given by
\be  m_N = {m_N^0}-{4c_1}m_\pi^2 -\frac{3 g_A^2 }{16\pi f_\pi^2} m_\pi^3 .\ee
We perform a fit  to the volume corrected results at $\beta=3.9$, $\beta=4.05$ and $\beta=4.2$  and extract  $r_0=0.462(5)$~fm.
 Fitting instead to
the  $\beta=3.9$, $\beta=4.05$ results we find  $r_0=0.465(6)$~fm showing
that indeed cut-off effects are small. To estimate
the error due to the chiral extrapolation we use HB$\chi$PT to ${\cal O}(p^4)$, which leads to
$r_0=0.489(11)$.
 We take the difference between the ${\cal O}(p^3)$
and  ${\cal O}(p^4)$ mean values 
as an estimate of the uncertainty due to the chiral extrapolation.
 Fits to other higher order $\chi$PT formulae shown in Fig.3 and 
described in Ref.~\cite{ETMC-baryons}
are consistent with  ${\cal O}(p^4)$ HB$\chi$PT.
Using $r_0=0.462(5)(27)$ and  the computed $r_0/a$ ratios we obtain
$ a_{\beta=3.9}=0.089(1)(5)$, $a_{\beta=4.05}=0.070(1)(4)$ and $a_{\beta=4.2}=0.056(2)(3)$.
 These values are consistent with the lattice spacings determined from 
$f_\pi$ and will be used
for converting to physical units in what follows.
 We note that results on the nucleon mass using twisted mass fermions agree with
those obtained using other lattice ${\cal O}(a^2)$ formulations
for lattice spacings below 0.1~fm~\cite{Alexandrou:2009}. \\
\\
\begin{minipage}{0.48\linewidth}
 \includegraphics[width=\linewidth]{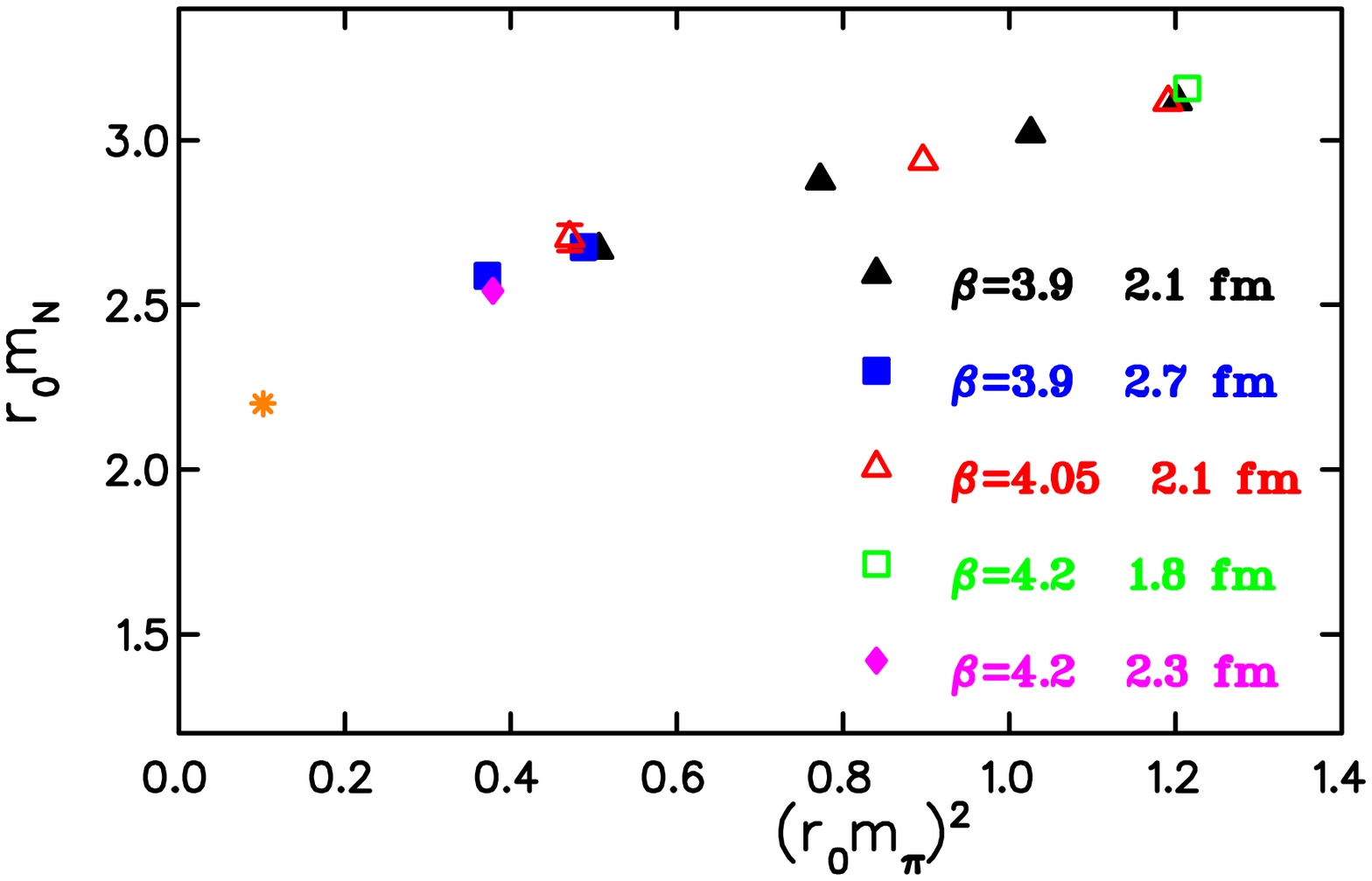}
{\small Fig.2: Nucleon mass in units of $r_0$ at 3 lattice spacings and spatial
lattice size $L$ such that  $m_\pi L> 3.5$.}
\end{minipage}\hfill
\begin{minipage}{0.48\linewidth}
   \includegraphics[width=\linewidth]{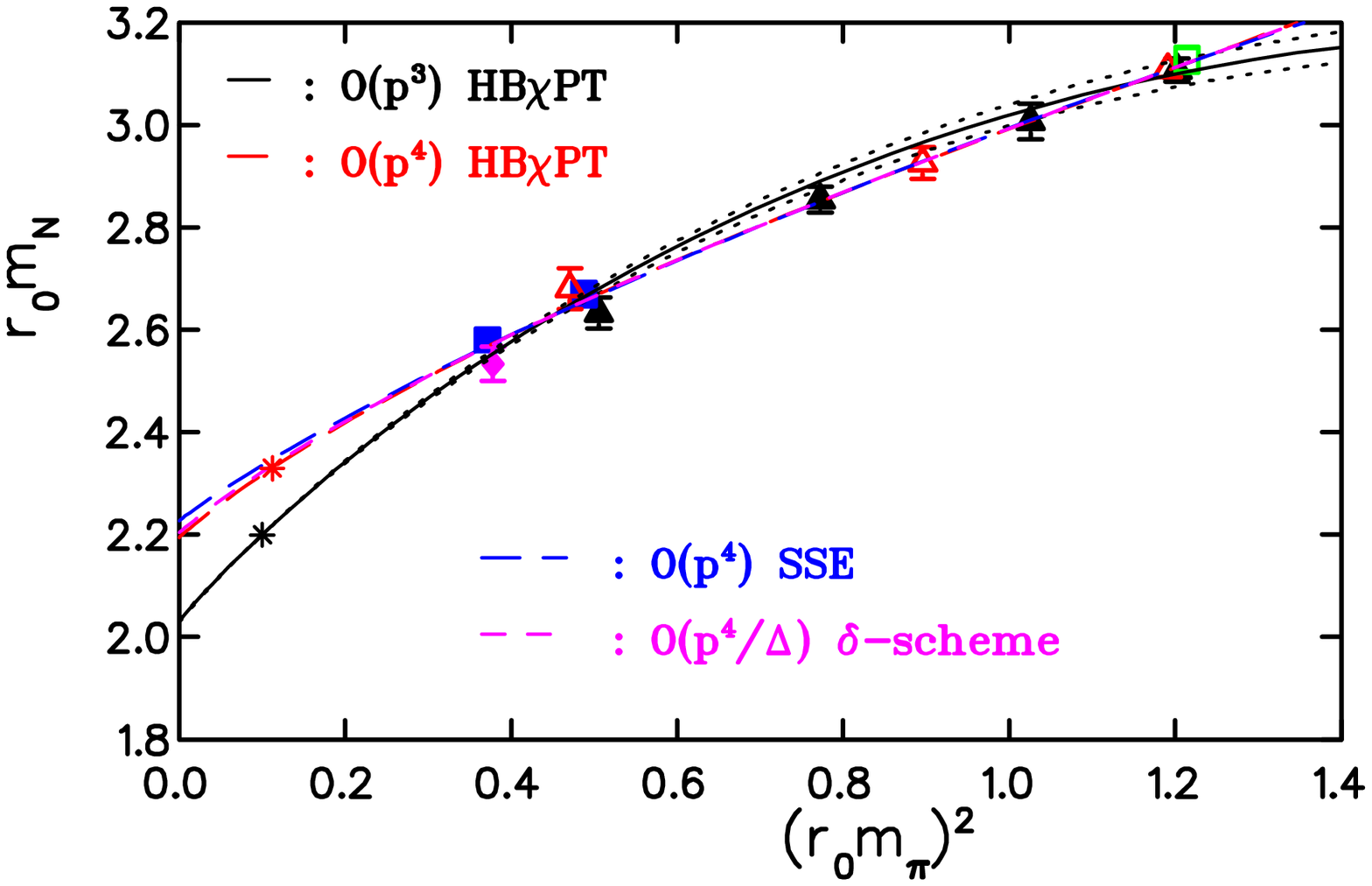}
{\small Fig.3: Nucleon mass in units  $r_0$. The solid and dashed lines are  fits to ${\cal O}(p^3)$ and ${\cal O}(p^4)$ HB$\chi$PT. }
\end{minipage}\vspace*{0.2cm}

In order to calculate the aforementioned nucleon matrix elements 
we calculate respectively
the two-point and three-point functions:
$G(\vec q, t_f) =\sum_{\vec x_f} \, e^{-i\vec x_f \cdot \vec q}\, 
     {\Gamma^0_{\beta\alpha}}\, \langle {J_{\alpha}(t_f,\vec x_f)}{\overline{J}_{\beta}(0)} \rangle $ and
 $ G^\mu(\Gamma^\nu,\vec q, t) =\sum_{\vec x, \vec x_f} \, e^{i\vec x \cdot \vec q}\,  \Gamma^\nu_{\beta\alpha}\, \langle {J_{\alpha}(t_f,\vec x_f)} j^\mu(t,\vec x) {\overline{J}_{\beta}(0)} \rangle
$, where the projection matrices
   ${\Gamma^0} = \frac{1}{4}(\eins + \gamma_0)$ and 
   ${\Gamma^k} = i{\Gamma^0} \gamma_5 \gamma_k$. 
The kinematical setup that we used is illustrated in Fig.~1: We create
the nucleon at $t_i=0$, $\vec x = 0$ (source) and annihilate it 
at $t_f/a=12$, $\vec p_f = 0$ (sink). We checked that the sink-source
time separation of $12a$ is sufficient for the isolation of the 
nucleon ground state by comparing the results with those obtained when
 $t_f/a=14$ is used~\cite{Alexandrou:2008rp}. We insert the current
 $j^\mu$ at $t$ carrying momentum $\vec q = -\vec p_i$.
In this work we limit ourselves to the  calculation of
the connected diagram which in the isospin limit yields the isovector
electromagnetic form factors. This is  calculated
 by performing sequential inversions through the sink so that
 no new inversions are needed for different operator $j^\mu(t,\vec q)$.
However new inversions are necessary for a different choice of the projection matrices $\Gamma^\alpha$. In this work, we consider the four choices given above, which
are optimal for the form factors considered here and construct the ratio
\be
R^{\mu} = \frac{G^\mu(\Gamma,\vec q,t) }{G(\vec 0, t_f)}\ \sqrt{\frac{G(\vec p_i, t_f-t)G(\vec 0,  t)G(\vec0,   t_f)}
                                                                                   {G(\vec 0  , t_f-t)G(\vec p_i,t)G(\vec p_i,t_f)}}
     \stackrel{t_f-t,t\rightarrow \infty}{\longrightarrow} \Pi^\mu (\Gamma,\vec q) \quad.
   \ee
The  leading time dependence and overlap factors cancel yielding 
as the plateau value
$ \Pi^\mu (\Gamma,\vec q) $ from which we extract the form factors using the relations
	    $$\Pi^\mu(\Gamma^0,\vec q) = \frac{c}{2m}[(m+E)\delta_{0,\mu} +  i q_k \delta_{k,\mu}]\ {\, G_E(Q^2)},\quad
	    \Pi^i(\Gamma^k,\vec q) 
= \frac{c}{2m}\sum_{jl}\epsilon_{jkl} q_j \delta_{l,i} {\, G_M(Q^2)} $$
	    ${\rm and}\quad \Pi^{5i}(\Gamma^k,\vec q) = \frac{ic}{4m}\left[\frac{q_k q_i}{2m}\ G_p(Q^2)-(E+m)\delta_{i,k}\ G_A(Q^2)\right],\quad k=1,\cdots,3, \quad{\rm where}\quad c=\sqrt{\frac{2m^2}{E(E+m)}}.$


\section{Results}
\subsection{Isovector electromagnetic form factors}
The  isovector electric and magnetic form factors at $\beta=3.9$ 
are shown in Figs.~4 and 5 respectively on lattices of size $24^3\times 48$ and
$32^3\times 64$ and for the pion masses and statistics listed in Fig.~6.\\
\\
\begin{minipage}{0.48\linewidth}
    \includegraphics[width=\linewidth]{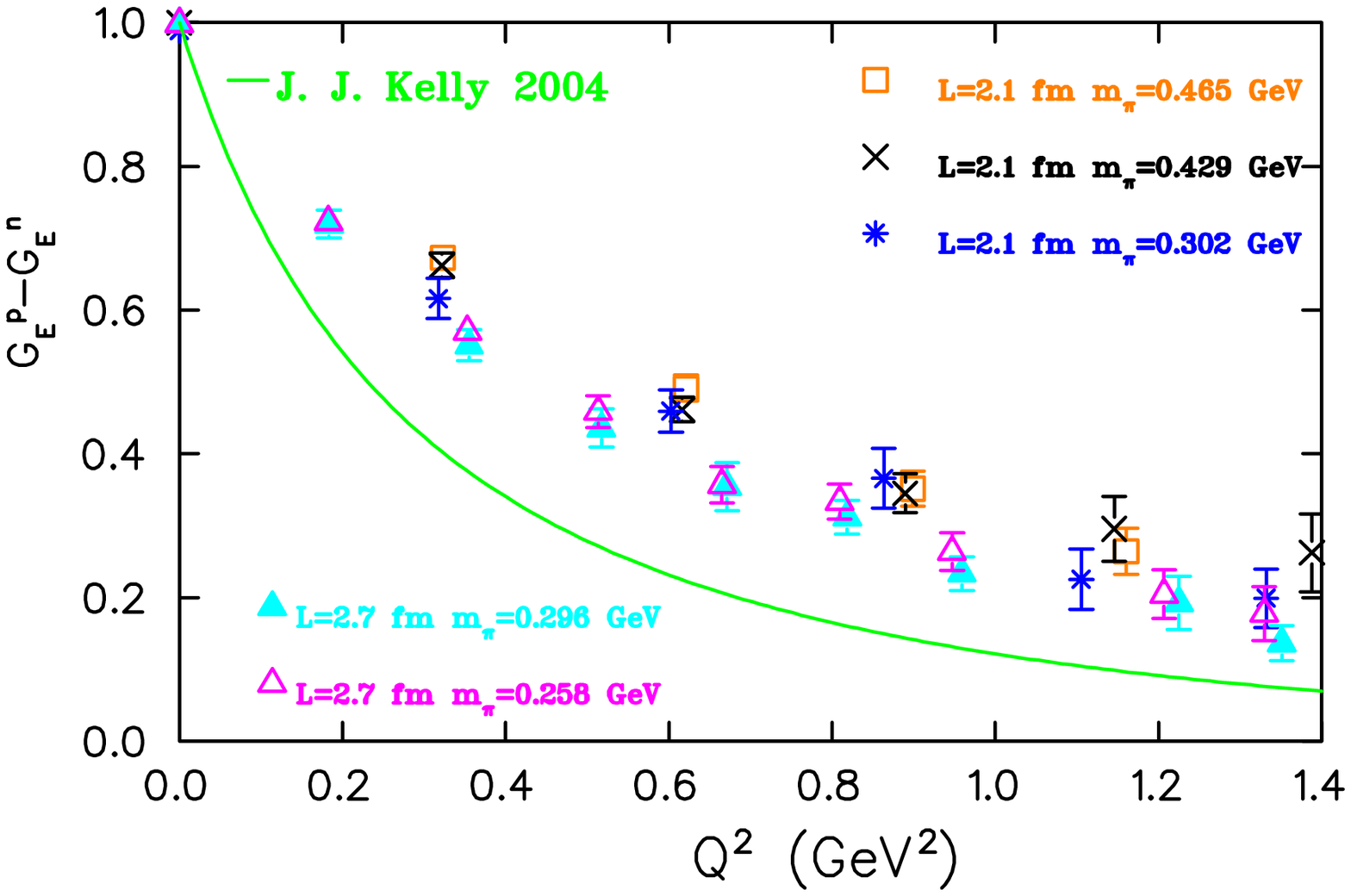} 
{\small Fig.~4: The isovector electric form factor as a function of $Q^2$ compared to experiment (solid curve).}
   \end{minipage}\hfill
\begin{minipage}{0.48\linewidth}
     \includegraphics[width=\linewidth]{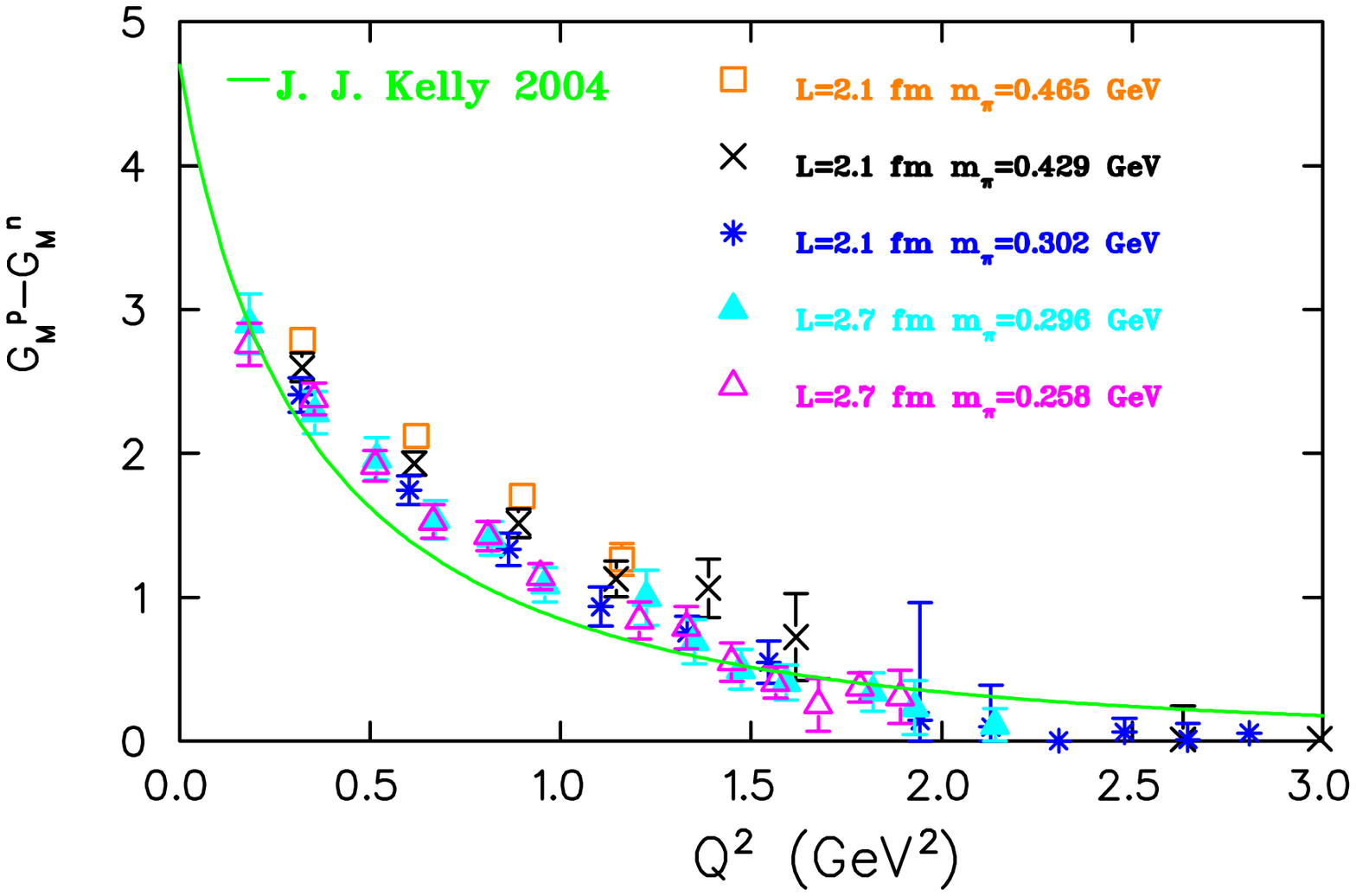}
{\small Fig.~5:The isovector magnetic form factor as a function of $Q^2$ compared to experiment (solid curve).}
   \end{minipage}

\vspace*{0.2cm}

By fitting the isovector magnetic form factor to a dipole form we extract
the isovector magnetic moment and  Dirac and Pauli root mean squared (r.m.s.) radii
shown in Figs.~6 and 7. As can be seen, the results obtained using
twisted mass fermions are in agreement with recent results using dynamical 
domain
wall fermions (DWF). 
Heavy baryon chiral perturbation theory to one-loop~\cite{QCDSF_chiral}
can be used to extrapolate to the physical point. In the case of the Dirac r.m.s radius we
fit the product $r_2^2\kappa_v$ so that only one fit parameter enters just as
in the case of $r_1^2$. This shifts the curve but does not affect its slope.
 We show fits
to our results alone as well as when we include the results obtained by the RBC-UKQCD
collaborations~\cite{Ohta:2008kd}.
The magnetic moment with three fit parameters is reproduced whereas  for $r_1^2$  we
obtained a weaker dependence   on the pion mass as compared to the one
predicted in chiral expansions.


\begin{minipage}{0.47\linewidth}
\begin{tabular}{ccc}
     $m_\pi$~(GeV) & no. of confs & L\\
      0.258 & 667 & 32\\
      0.296 &231 & 32 \\
      0.302 & 944 & 24 \\
      0.373 &210  & 24 \\
      0.429 &365 & 24 \\
      0.465 & 477 & 24\\
\end{tabular}
\hspace*{-0.7cm}\includegraphics[width=\linewidth]{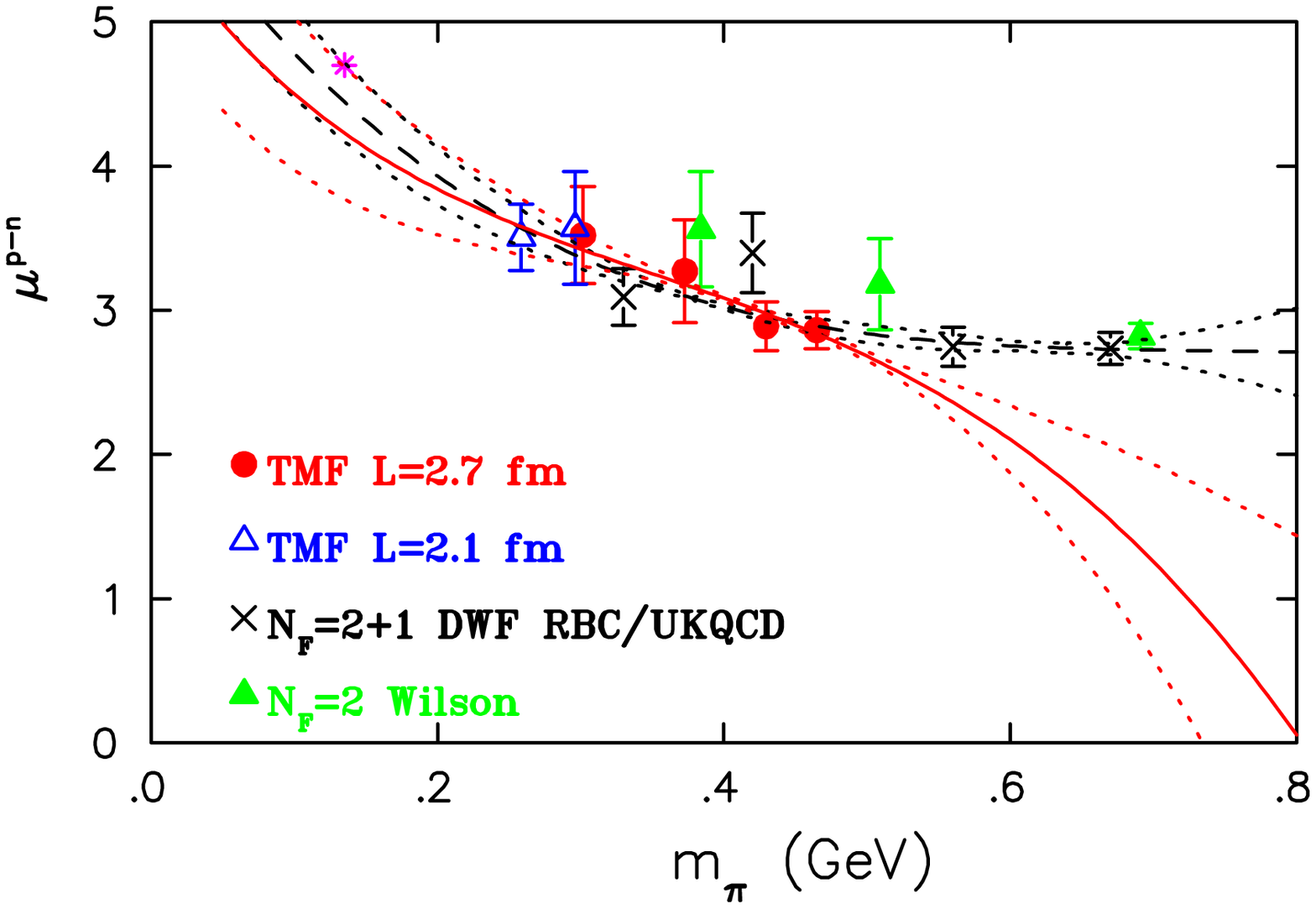}
{\small Fig. 6: The anomalous isovector magnetic moment of the nucleon.}
\end{minipage}\hfill
\begin{minipage}{0.47\linewidth}
\includegraphics[width=\linewidth]{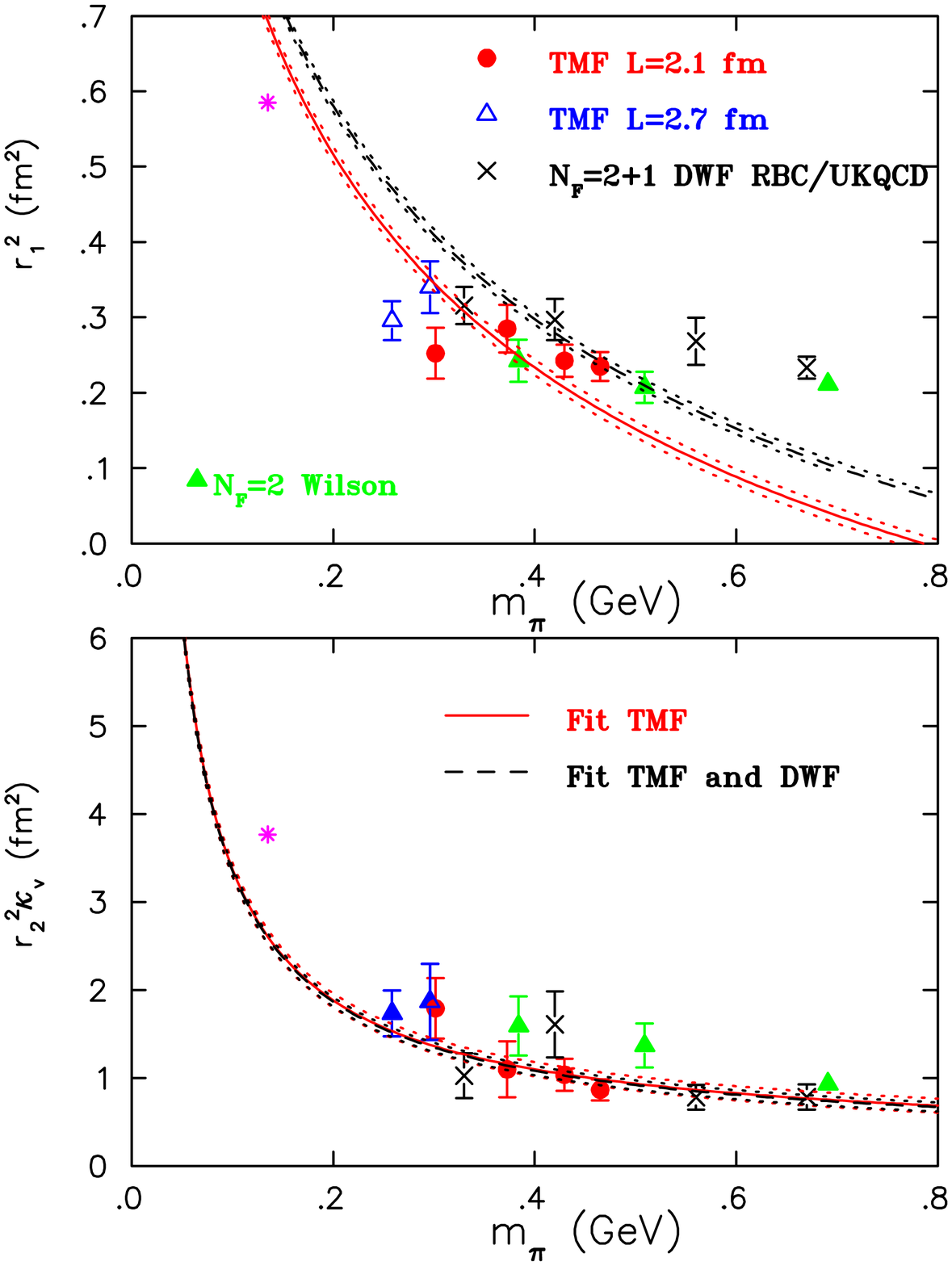}\vspace*{-0.2cm}
{\small Fig 7: The Dirac (upper) and Pauli (lower)  r.m.s radii.}
\end{minipage}



\subsection{Axial charge}
Our results on the nucleon axial charge are shown in Fig.~8 and are in
agreement with those obtained using domain wall fermions. Within our errors
no sizable finite volume effects are observed. In order
to extrapolate to the physical  point we use
 one-loop chiral perturbation theory  in the small scale expansion~\cite{ga-chiral}.\\
\begin{minipage}{0.37\linewidth}
There are three parameters to
fit: $g_A(0)$, the value of the axial charge at the chiral point, 
$g_1$ and a counter-term  $C_{SSE}$.
 Fitting using the TMF results we find $g_A(0)=1.10(13)$, $g_1=6.05(4.66)$ and 
$C_{SSE}=-3.65(3.58)$. The parameters $g_1$ and $C_{SSE}$
are highly correlated explaining the resulting large error band. 
Including  in the fit  the DWF data that span  larger pion masses 
 we obtain a very different curve, showing the sensitivity in the chiral
extrapolation of $g_A$. 
\end{minipage}\hfill
   \begin{minipage}{0.6\linewidth}
      \includegraphics[width=\linewidth]{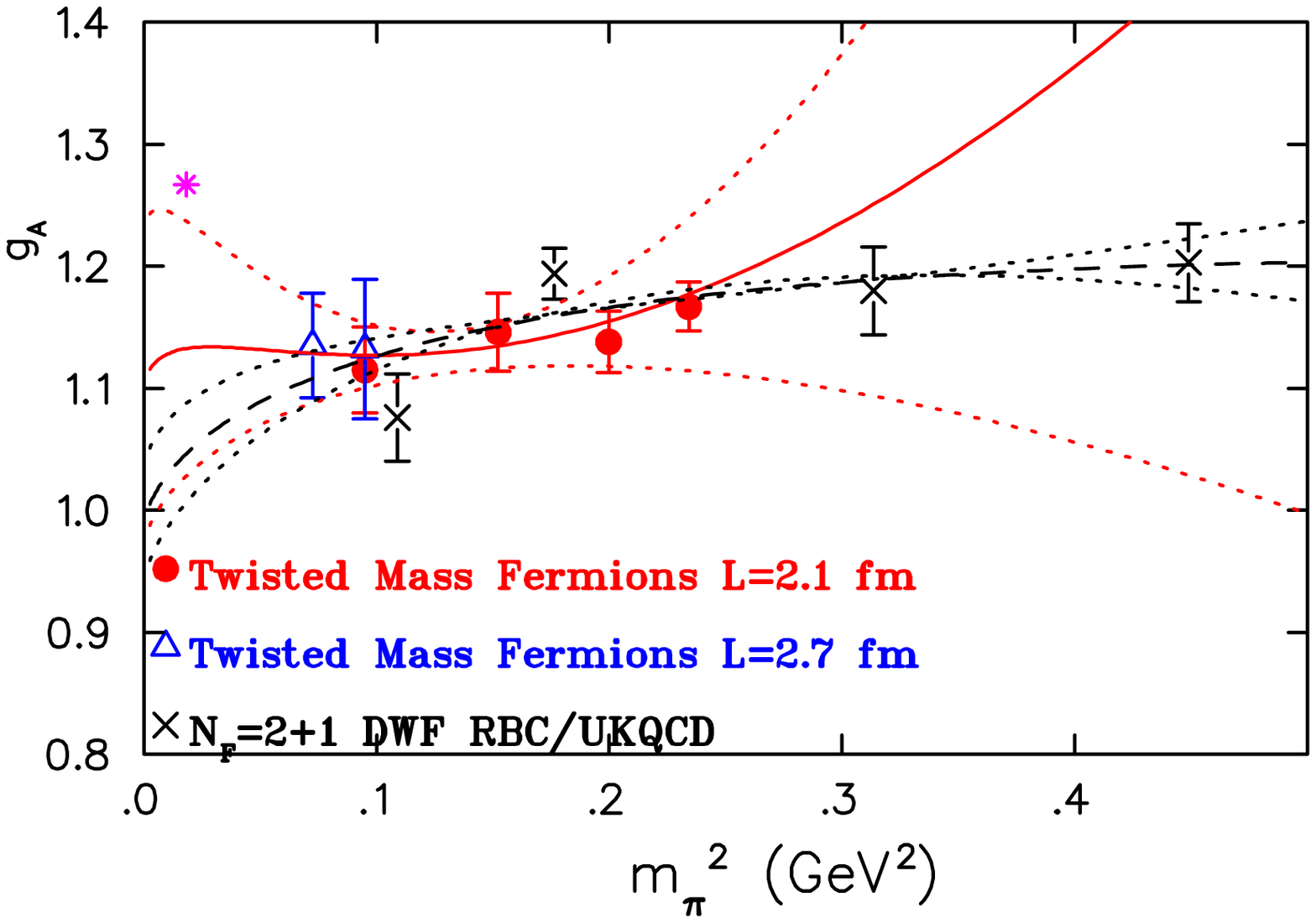}
{\small Fig.~8:  The nucleon axial charge. The solid (dashed) curve is
a chiral fit using TMF (TMF and DWF) results.}
   \end{minipage}

\subsection{Axial form factors}
Results for the axial form factors $G_A(Q^2)$ and $G_p(Q^2)$
are shown in  Figs.~9 and 10 respectively. We perform a
dipole fit to $G_A(Q^2)$ 
 extracting
an axial mass larger than in  experiment as expected from the smaller slope
shown by the lattice data both for TMF and DWF.
Assuming pion pole dominance we can
relate the form factor $G_p(q^2)$ to $G_A(Q^2)$. Using the pion mass measured
on the lattice we predict the dashed curve shown in Fig.~10. Our lattice
data on $G_p(q^2)$ are flatter than pion pole dominance predicts requiring
a larger pole mass than the pion mass.
Large volume effects are expected at low $Q^2$ indicated by the deviation
 of $G_p(Q^2)$ from the fitted curve at the smallest $Q^2$-value. \\
 \begin{minipage}{0.49\linewidth}
\includegraphics[width=\linewidth]{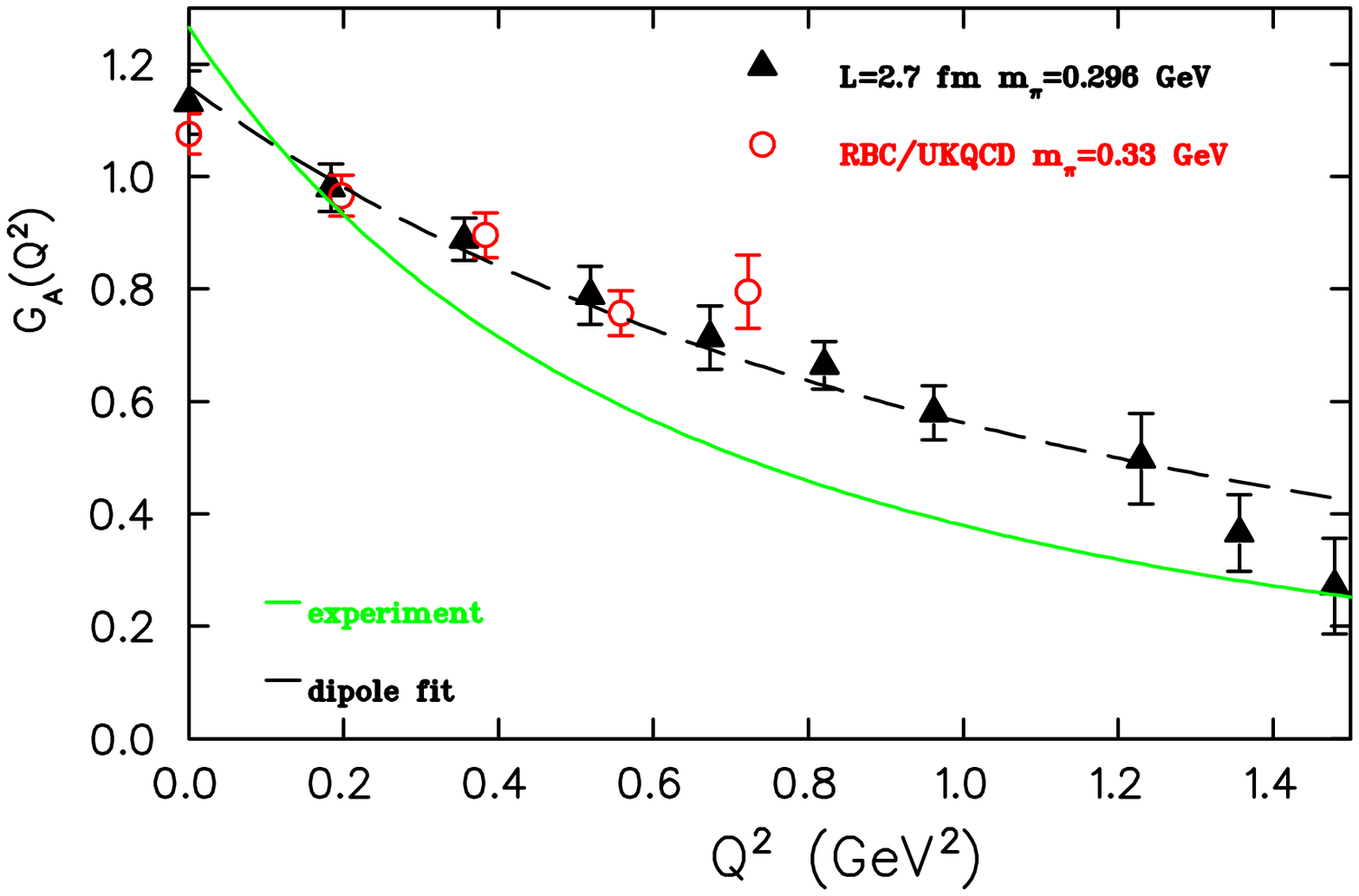}            
{\small Fig. 9: Axial form factor $G_A(Q^2)$ as a function of $Q^2$. 
The dashed (solid) line
is the best dipole fit to the lattice (experimental) results.}
\end{minipage}\hfill 
\begin{minipage}{0.49\linewidth}
\includegraphics[width=\linewidth]{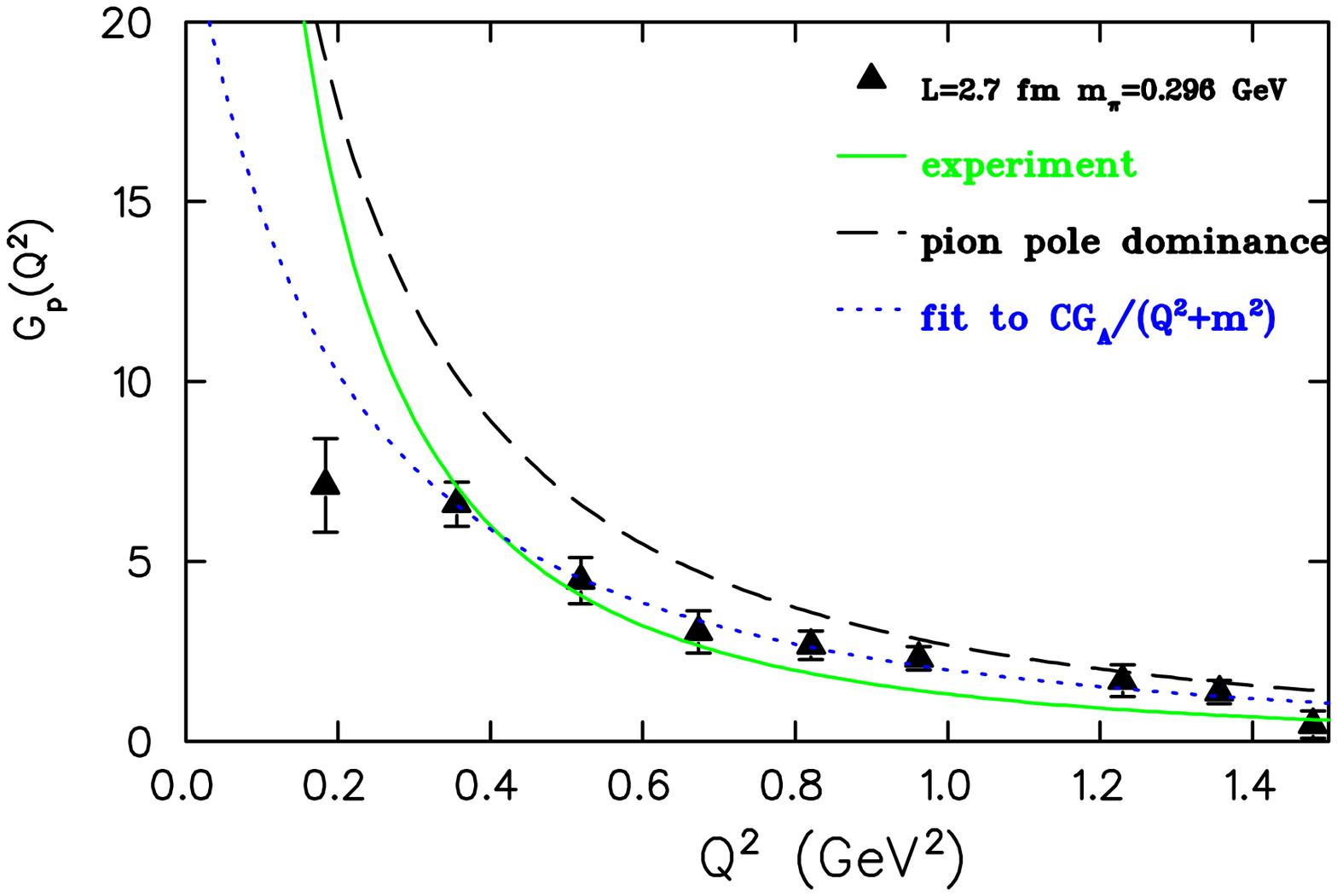}
{\small Fig. 10: The dotted line is a fit of lattice
results to the form $\frac{CG_A(Q^2)}{(Q^2+m^2)}$ and the dashed (solid) line is the prediction of pion pole dominance using lattice (experimental) results on $G_A(Q^2)$.} 
   \end{minipage}

\section{Conclusions}
Using $N_F=2$ twisted mass fermions we obtain
accurate results on the isovector electromagnetic 
$G_E, G_M$ and  axial $G_A, G_p$ form factors as a function
of $Q^2$  for pion
mass in the range of about  260-470 MeV. The general feature is a flatter
dependence on $Q^2$ than experiment. The  Dirac r.m.s radius thus
shows a weaker dependence on the pion mass than expected from chiral perturbation theory. Finite volume effects
are found to be small on quantities like  $g_A$ and the isovector magnetic moment and r.m.s radii.	 Our results are in agreement with recent results obtained using dynamical $N_F=2+1$ DWF. 
At the physical point using TMF we find  $g_A= 1.13(10)$ close to the 
experimental value albeit with a large error due to the chiral extrapolation.
An analysis of these form factors 
  at $\beta=4.05$ is under way so that a check of cut-off effects
can be carried out.

\vspace*{0.3cm}
\noindent
 {\bf Acknowledgments:} This work was performed using HPC resources from GENCI (IDRIS and CINES) Grant 2009-052271
and was partly supported by funding received   by  the DFG
Sonder\-for\-schungs\-be\-reich/ Trans\-region SFB/TR9 and
the
 Cyprus Research Promotion Foundation under contracts EPYAN/0506/08,
KY-$\Gamma$/0907/11/ and  TECHNOLOGY/$\Theta$E$\Pi$I$\Sigma$/0308(BE)/17.

\end{document}